\documentclass[runningheads]{llncs}
\usepackage[T1]{fontenc}
\usepackage{graphicx}
\usepackage{enumitem}
\usepackage{multirow}%
\usepackage{amsmath,amssymb,amsfonts}%
\usepackage{mathrsfs}%
\usepackage[figuresright]{rotating}%
\usepackage{xcolor}%
\usepackage{textcomp}%
\usepackage{caption}
\usepackage{subcaption}
\usepackage{booktabs}
\usepackage{manyfoot}%
\begin{document}
\title{Influence of the microstructure on the mechanical behavior of nanoporous materials under large strains}
\titlerunning{Influence of the microstructure of nanoporous materials}

%

\author{Rajesh Chandrasekaran\inst{1} \and
Mikhail Itskov\inst{1} \and
Ameya Rege \inst{2,3}}
\authorrunning{R. Chandrasekaran et al.}
%
\institute{Department of Continuum Mechanics, RWTH Aachen University, Eilfschornsteinstr. 18, 52062 Aachen, Germany \and Department of Aerogels and Aerogel Composites, Institute of Materials Research, German Aerospace Center, Linder Höhe, 51147 Cologne, Germany \and School of Computer Science and Mathematics, Keele University, Staffordshire ST5 5B4, United Kingdom}
\maketitle              
\begin{abstract}
Nanoporous materials are characterized by their complex porous morphology illustrated by the presence of a solid network and voids. The fraction of these voids is characterized by the porosity of the structure, which influences the bulk mechanical properties of the material. Most literature on the mechanics of porous materials has focused on the density-dependence of their elastic properties. In addition to porosity, other pore characteristics, namely pore-size and shape described by the pore-size distribution, and pore-wall size and shape, also influence the bulk response of these materials. In this work, the mechanical structure-property relation of nanoporous materials is studied under large deformations using a computational framework. The interdependent microstructural parameters are identified. After a successful correlation between the synthesis and microstructural parameters, the synthesis of porous materials can be guided and optimized by controlling these parameters.

\keywords{Nanoporous materials \and Structure-property relation \and Pore-size distribution \and Porosity \and Large deformation \and Inelastic properties.}
\end{abstract}
\section{Introduction}\label{sec1}
Due to depleting resources, rising cost of raw materials, and advancement in sustainability, reduced material consumption in automobiles, aerospace and machinery has become a continued challenge. Therefore, there is an ever increasing demand for the use of lighter and multifunctional innovative materials with improved mechanical and thermal properties. One of the promising class of materials for light-weight design and multifunctional applications are porous materials, which consist of an interconnected solid structure around a porous space across scales resulting in relatively high structural rigidity and low mass density. Nanoporous materials belong to a special class of porous solids with pore-sizes comparable to 100 nm or smaller. Such materials exhibit fascinating mechanical and thermal properties, namely ultralow density, high energy absorption capacity, better thermal management and durability under dynamic loads \cite{bib1,bib2}. Typical examples of natural and synthetic nanoporous solids are zeolites, activated carbon, metal-organic frameworks, ceramics, silicates, nonsiliceous materials, aerogels, various polymers, and inorganic porous hybrid materials \cite{bib1}.

In the last few decades, research on the synthesis, characterization, functionalization, molecular modeling, and design of new and novel nanoporous materials have witnessed an exponential rise due to their increasing demand in a wide spectrum of applications. Porous materials can be synthesized using different fabrication techniques such as precipitation, solid-state reaction (usually performed at high temperature), sol-gel, hydrothermal, and solvothermal synthesis routes \cite{bib3,bib4,bib5}. Exploring diverse materials to form the skeletal backbone for designing the solid network with different pore morphologies, enables us to achieve a unique and attractive combination of their mechanical and thermal properties. The structure, pore-size and subsequently the pore fraction, specific surface area, and density of nanoporous materials could be tailored using different fabrication techniques. With this objective, characterizing the structure-property relationship of such materials is key to their material design. They are typically characterized by two essential microstructural features, namely density and pore-size distribution (PSD). Density can be interpreted as relative density which can be expressed in terms of porosity. Porosity can be defined as $\phi = 1 - \phi_s$, where $\phi_s$ is the solid fraction of the structure, i.e., the fraction of total volume taken up by the solid phase. PSD characterizes the relative abundance of each pore-size in a representative volume of the material. Both microstructural parameters can be measured experimentally using different characterization methods.


Nanoporous materials are widely used in diverse applications, ranging from energy absorbers in crash applications, scaffolds in tissue engineering, carrier materials for drugs and food packaging, and all applications that demand a stable mechanical performance with insulation characteristics. Some of these materials show brittle behavior under tensile loading and undergo elasto-plastic deformation under large strain when subjected to compression. On the other hand, there also exist highly flexible materials. Most tests on cellular-like nanoporous materials are carried out under compressive loading where significant energy absorption can be observed. Under monotonic loading, the typical behavior of porous material is characterised by three major phenomena that give rise to the consequent defomation zones, respectively: (a) a small elastic regime where the pore walls undergo buckling or bending, followed by (b) a plateau regime which arises from plastic yielding and pore collapse, and (c) the densification regime (resulting in a rapid rise of peak stress with a considerably small increase in strain \cite{bib8}), where the collapsed pores begin to densify resulting in hardening of the network. Under cyclic loading, nanoporous materials show inelastic behavior with a hysteresis cycle and large residual deformations \cite{bib16}.

The main challenge remains the correlation of the structural features with the properties, thus enabling reverse engineering through optimization techniques. In comparison to trial and error-lab-based experiments, numerical modeling and computer simulation is a cost- and resource- efficient method to study such structure-property relations. Gibson and Ashby \cite{bib9} expressed the mechanical properties such as Young's modulus and compressive strength as power laws with respect to the densities in porous materials. Also, the micromechanical model recently developed by Rege et al.~\cite{bib11,bib18} has shown good predictive capabilities in describing the mechanical behavior of aerogel materials. However, these models only account for the PSD but disregards the heterogeneous morphology of the pore structure in terms of its shape. The PSD describes the spatial variation of the pore-sizes and has recently been shown to influence the mechanical properties of porous materials \cite{bib11,bib12}. These nanoporous materials can be computationally designed by models which inherit the most essential geometric properties, such as the solid fraction and PSD. The most widely used computational methods for the reconstruction of porous structure are Voronoi tessellations. Recently, Chandrasekaran et al. \cite{bib13} reconstructed the 3-d nanoporous microstructure of biopolymer aerogels using Laguerre-Voronoi tesselation (LVT) based on random closed packing of polydisperse spheres (RCPPS). This model approach requires PSD and solid fraction to adequately represent the 3-d nanoporous morphology of the material, while still proving to be accurate in predicting the bulk macroscopic behavior.

The present work aims at providing a better understanding of the structure-property relation to guide and optimize the synthesis of nanoporous material. This can be achieved by correlating synthesis characteristics to model parameter as shown in \cite{bib10}. With this as the major goal, the influence of different geometric parameters on the bulk mechanical response is computationally studied using the framework proposed in \cite{bib13}. To this end, the elastic and inelastic response of various nanoporous structures with PSD based on different probability density functions (PDFs) and different combinations of other geometric properties is studied under large compressive deformations. Geometric properties which are interdependent and more sensitive to the bulk response are identified.

\section{Methods}\label{sec2}
In this section, the method of determining the PSD for any given choice of PDF, namely beta, log-normal or normal distributions is discussed. In addition, the overview of geometric and finite element (FE) modeling based on the given probability is elaborated.

Generally, irrespective of the experimental methodology, the PSD is derived by step-wise determination of pore volume increments for corresponding pore-width intervals. Accordingly, the sum of all the pore volume increments represents the total pore volume. Since for the following study, spherical pores are assumed, the pore-width can be expressed by pore diameter $D$. Note that the spherical pore shape is only an initial assumption and therefore modeled structure exhibits a random pore shape. The PSD is represented by a function $P(D)$ proportional to the combined volume of all pores whose effective pore diameter is within an infinitesimal range centered on $D$. Hence, the PSD can be obtained from the ratio of the pore volume increment $\Delta V_i$ for each pore-width $D_i$ and the total pore volume, written as
\begin{equation}
P(D_i) = \frac{\Delta V_i }{\sum\limits_{j=1}^{n} \Delta V_j},
\end{equation}

\noindent
where $n$ is the total number of pore-width intervals.

The cumulative pore-size distribution $P_c$ can be obtained from the ratio of the partial sums of the pore volume increment, i.e., that of the cumulative pore volume to the total pore volume, written as

\begin{equation}
P_c(D_i) = \frac{\sum\limits_{k=1}^{i}  \Delta V_k }{\sum\limits_{j=1}^{n} \Delta V_j}.
\end{equation}

\subsection{Determination of pore-size distribution (PSD) from probability density function (PDF)}\label{sec:2.2}
The equivalent PSD is determined from the PDF for a given set of pore-diameter intervals $D$ and binwidth $dD$. The set of pore-diameter intervals is represented as $D= \{D_i\}$ with $i = 1, 2, 3,\dots, n$.

The volume of each pore with diameter $D_i$ can be expressed by
\begin{equation}
V_i = f_1(D_i). \label{eq:3}
\end{equation}

Further, the number of pores having a diameter between $D_i$ and $D_i+dD$ takes the form 
\begin{equation}
dN_i = N f_2(D_i) dD. \label{eq:4}
\end{equation}
where $f_2$ is the probability density function (PDF) and $N$ is the total number of pores in a system.

The volume of all pores with diameter between $D_i$ and $D_i+dD$ can thus be written as 
\begin{equation}
dV_i = V_idN_i = N f_1(D_i) f_2(D_i) dD. \label{eq:5}
\end{equation}

The theoretical pore-size distribution can be obtained from the ratio of the volume of all pores with diameters between $D_i$ and $D_i+dD$ to the total pore volume in the system, 
\begin{equation}
P(D_i) = \frac{ dV_i}{\sum\limits_{j=1}^{N} dV_j} =\frac{ f_1(D_i)f_2(D_i)}{\sum\limits_{j=1}^{N} f_1(D_j)f_2(D_j)}. \label{eq:6}
\end{equation}

\begin{figure}[hb!]
    \centering 
    \includegraphics[width=0.9\linewidth]{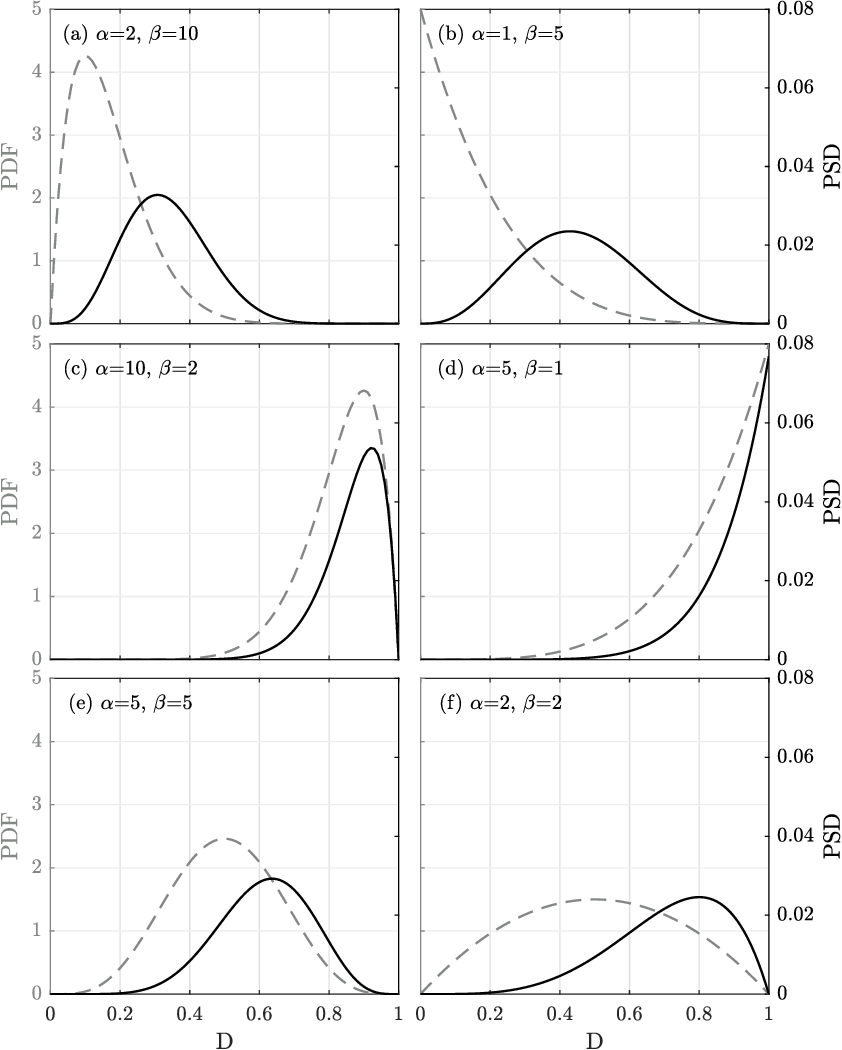}
    \caption{Comparison between the PDF and PSD for beta distribution with different skewness: (a) \& (b) Right-skewed ($\alpha < \beta$), (c) \& (d) Left-skewed ($\alpha > \beta$), and (e) \& (f) Symmetric ($\alpha = \beta$).}
    \label{fig:PDFvsPSD}
\end{figure}

The theoretical cumulative pore-size distribution can be obtained from the ratio of the sum of the volumes of all pores with diameter up to $D_i$ to the total volume of all pores in the system, 
\begin{equation}
P_c(D_i) =  \frac{\sum\limits_{k=1}^{i} dV_k}{\sum\limits_{j=1}^{n} dV_j}= \frac{\sum\limits_{k=1}^{i} f_1(D_i)f_2(D_i)}{\sum\limits_{j=1}^{n} f_1(D_j)f_2(D_j)}. \label{eq:7}
\end{equation}

In order to solve Eq. \ref{eq:3} \& \ref{eq:4}, it is necessary to obtain the explicit functions $f_1$ and $f_2$, which depend on the geometrical type of the pore model as follows.

\begin{itemize}
    \item For a spherical type pore model, $f_1(d) = \frac{\pi}{6} d^3$, where $d$ is the diameter of the spherical pore. 
    \item For a cylindrical type pore model, $f_1(d) = \frac{\pi}{4} l d^2$, where $d$ is the diameter of the and $l$ is the length of the cylinderical pore. 

    \item For the beta distribution, $f_2(x) = \frac{\Gamma (\alpha + \beta)}{\Gamma (\alpha) \Gamma (\beta)} \cdot x^{\alpha - 1} \cdot (1 - x)^{\beta - 1}$, where 

    \begin{itemize} [label={}]
        \item[-] $x$ is the pore diameter in interval $[0, 1]$,
        \item[-] $\alpha$ and $\beta$ are the shape parameters,
        \item[-] $\Gamma$ is a gamma function.
    \end{itemize}

    \item For the normal distribution, $f(x) = \frac{1}{s \sqrt{2\pi} } e^{-\frac{1}{2}\left(\frac{x-\mu}{s}\right)^2}$, where 

    \begin{itemize} [label={}]
        \item[-] $x$ is the pore diameter, i.e., $x\in \mathbb{R} $,
        \item[-] $\mu$ is the mean,
        \item[-] $s$ is the standard deviation.
    \end{itemize}

        \item For the log-normal distribution, $f(x) = \frac{1}{x s \sqrt{2\pi} } e^{-\frac{1}{2}\left(\frac{ln x-\mu}{s}\right)^2}$, where 

    \begin{itemize} [label={}]
        \item[-] $x$ is the pore diameter in the interval $[0, +\infty]$ whose logarithm is normally distributed, 
        \item[-] $\mu$ is the mean of the natural logarithm of the variable in the interval $[-\infty, +\infty]$,
        \item[-] $s$ standard deviation of the natural logarithm of the variable, i.e., $s > 0$.
    \end{itemize}
\end{itemize}

The comparison between the probability density function and the corresponding pore-size distribution is illustrated in Fig. \ref{fig:PDFvsPSD} using a beta distribution for different shape parameters $\alpha$ and $\beta$ and assuming the pore shape to be spherical. It is observed that the contribution of the larger pores is dominant in the PSD, although the probability of occurrence of the smaller pores is higher. In  Fig. \ref{fig:PDFvsPSD}, a right-skewed PDF shows a symmetric PSD, and a symmetric PDF shows a left-skewed PSD. Thus, one should not misinterpret PSD as PDF and vice versa. This may result in totally different analysis and interpretation.

\subsection{Modeling of the open-porous network}\label{sec:2.3}
This section describes the geometric and finite element (FE) modeling of the porous structure using symmetric beta PDF with  $\alpha =5~\&~\beta = 5$ (henceforth referred to as reference PDF). A computational model of a porous structure with highly dispersed pore-sizes can be generated using a random closed packing of polydisperse spheres (RCPPS) and Laguerre-Voronoi tesselation (LVT). This modeling technique was shown to be a powerful method to generate and model porous materials computationally \cite{bib13}.

\subsubsection{Geometric modeling:}\label{sec:2.3.1}
In the modeling of porous structures, one has to account for the following geometric properties: porosity/solid fraction/relative density and PSD, which are the essential input parameters for the generation of the microstructure. In general, Voronoi tessellation is capable of representing the microstructure of porous materials with interconnected network structure. However, it limits the control over the PSD due to the randomized spatial distribution of seed points in 3-d space and the generation of cell boundaries equidistant between seed points. Therefore, obtaining a structure for a given PSD is a challenge using the classical Voronoi tessellation. LVT is a weighted version of Voronoi tessellations, where the pore-sizes can be controlled by defining specific weights for each seed. The resulting position of the seed in the 3-d space and the corresponding weights necessary for LVT are provided by RCPPS as sphere centers and radii respectively. Consequently, a periodic 3-d structure capturing a particular PSD with an interconnected solid network is generated. In addition to the PSD, a microstructure model with a given solid fraction can be obtained by defining appropriate cross-sectional properties to the cell walls of the Voronoi structure.

 \begin{figure}[ht!]
    \centering 
    \includegraphics[width=0.99\linewidth]{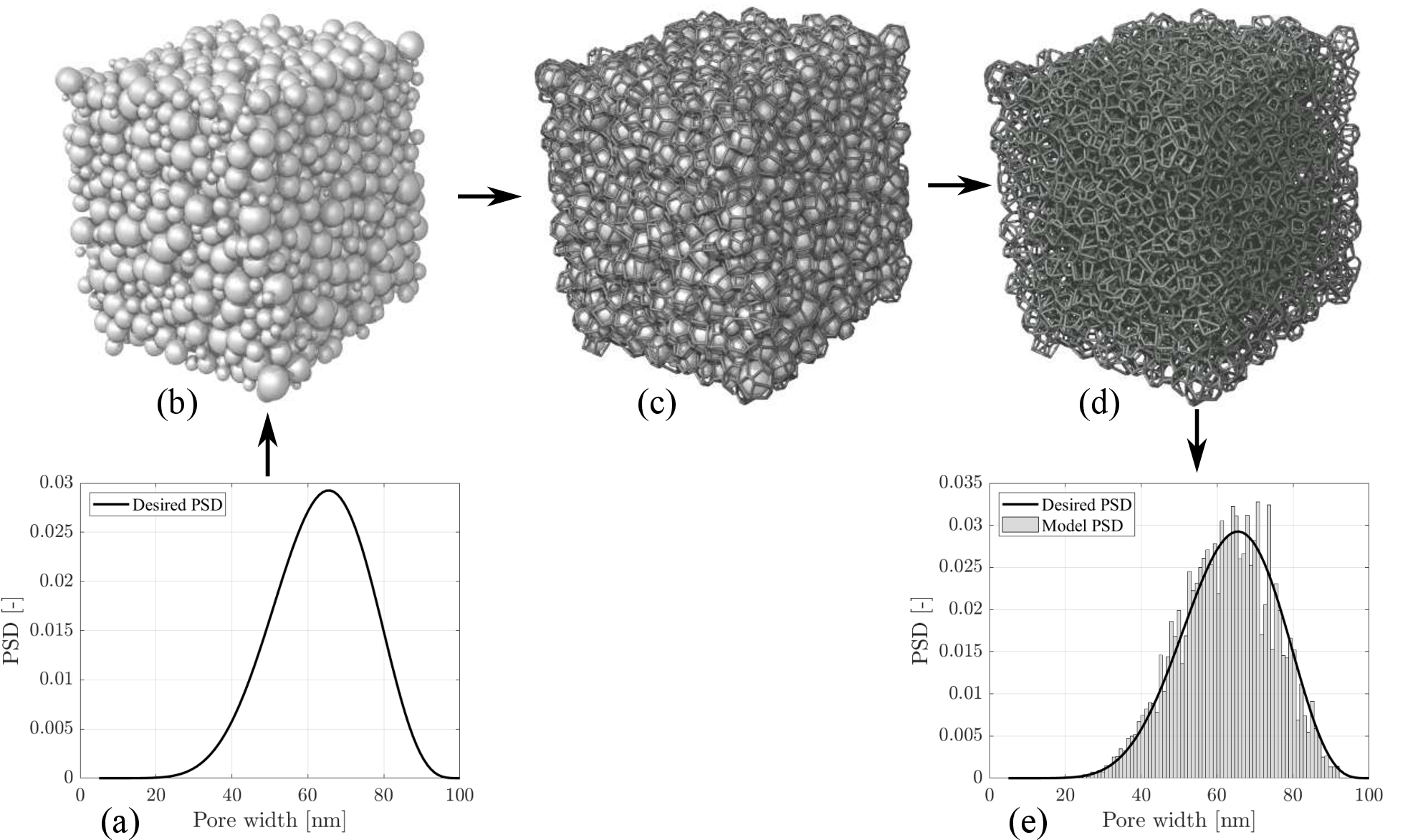}
    \caption{ Model workflow using RCPPS and LVT: (a) Desired PSD corresponding to reference PDF for N=3000 (b) RCPPS, (c) LVT, (d) Microstructure model, and (e) Comparison of desired PSD with the PSD of resulting microstructure model.}
    \label{fig:modelflow}
\end{figure}

In this study, we assume the cell walls to be cylindrical and therefore an appropriate cell wall diameter (CWD) has to be determined to obtain the microstructure model with the required solid fraction. This can be achieved by using two approaches as follows. (1) \textbf{Constant diameter (CD) approach}: a constant diameter to all the cell walls in the structure is determined by achieving a constant porosity to the overall structure. (2) \textbf{Diameter distribution (DD) approach}: a distinct diameter for each cell wall is determined by defining a constant porosity for each pore within the structure. The former method provides a structure with distinct porosity for each pore but has a constant CWD throughout, whereas the latter method provides a structure with constant porosity to each pore, but a distinct CWD. The latter method of determining the edge diameter is described in \cite{bib13}.
 
\begin{figure}[ht!]
    \centering 
    \includegraphics[width=0.6\linewidth]{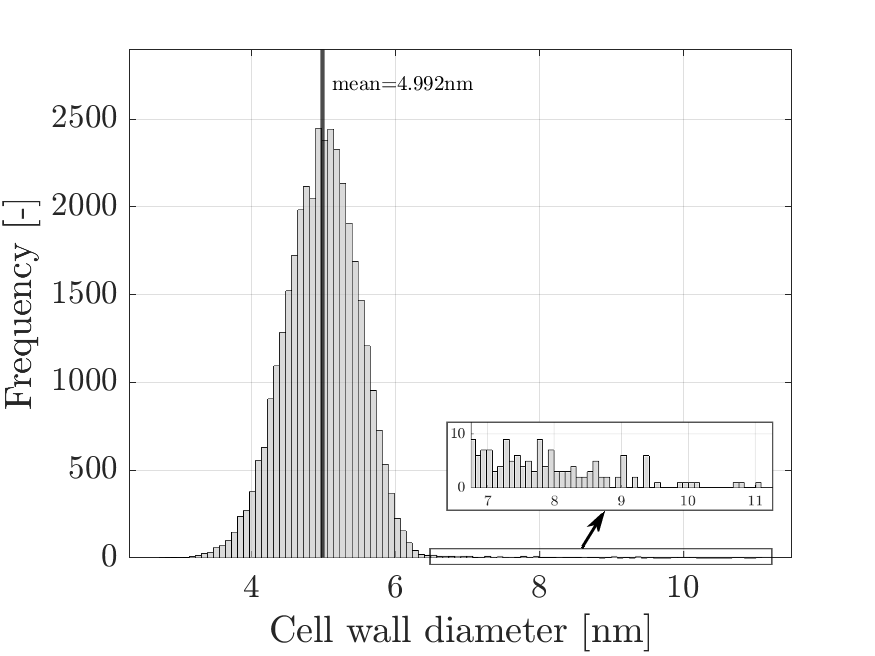}
    \caption{ CWD for a given solid fraction of 0.05, illustrating both approaches. Using CD approach with mean 4.992 nm, while using DD approach with histogram from 1.8 nm to 9.2 nm}
    \label{fig:walldia_hist}
\end{figure}

Generation of a computational microstructure model with pore-width range between 5-100 nm and $95\%$ porosity (solid fraction of 0.05) for a given PDF described by a symmetric beta distribution is illustrated in Fig.~\ref{fig:modelflow}. Based on the desired PSD (Fig.~\ref{fig:modelflow}a) and total number of pores (N) in the porous system, the size of the packing domain is determined and given as an input to the sphere packing algorithm. For each sphere of the closely packed set, the coordinates of the center and the corresponding radius are given as an input to the LVT algorithm (Fig.~\ref{fig:modelflow}b). LVT based on RCPPS is illustrated in Fig.~\ref{fig:modelflow}c. The resulting porous structure with desired microstructural properties is shown in Fig.~\ref{fig:modelflow}d. Its PSD can be compared with the desired input PSD shown as histogram in Fig.~\ref{fig:modelflow}e. For a given solid fraction of 0.05, the distribution of CWD of the structure obtained using DD approach and the mean diameter calculated using CD approach can be seen in Fig.~\ref{fig:walldia_hist}.

\subsubsection{FE modeling:}
\begin{figure*}[h!]
\centering
\begin{subfigure}[b]{0.52\textwidth}
  \centering
  \includegraphics[width=1\linewidth]{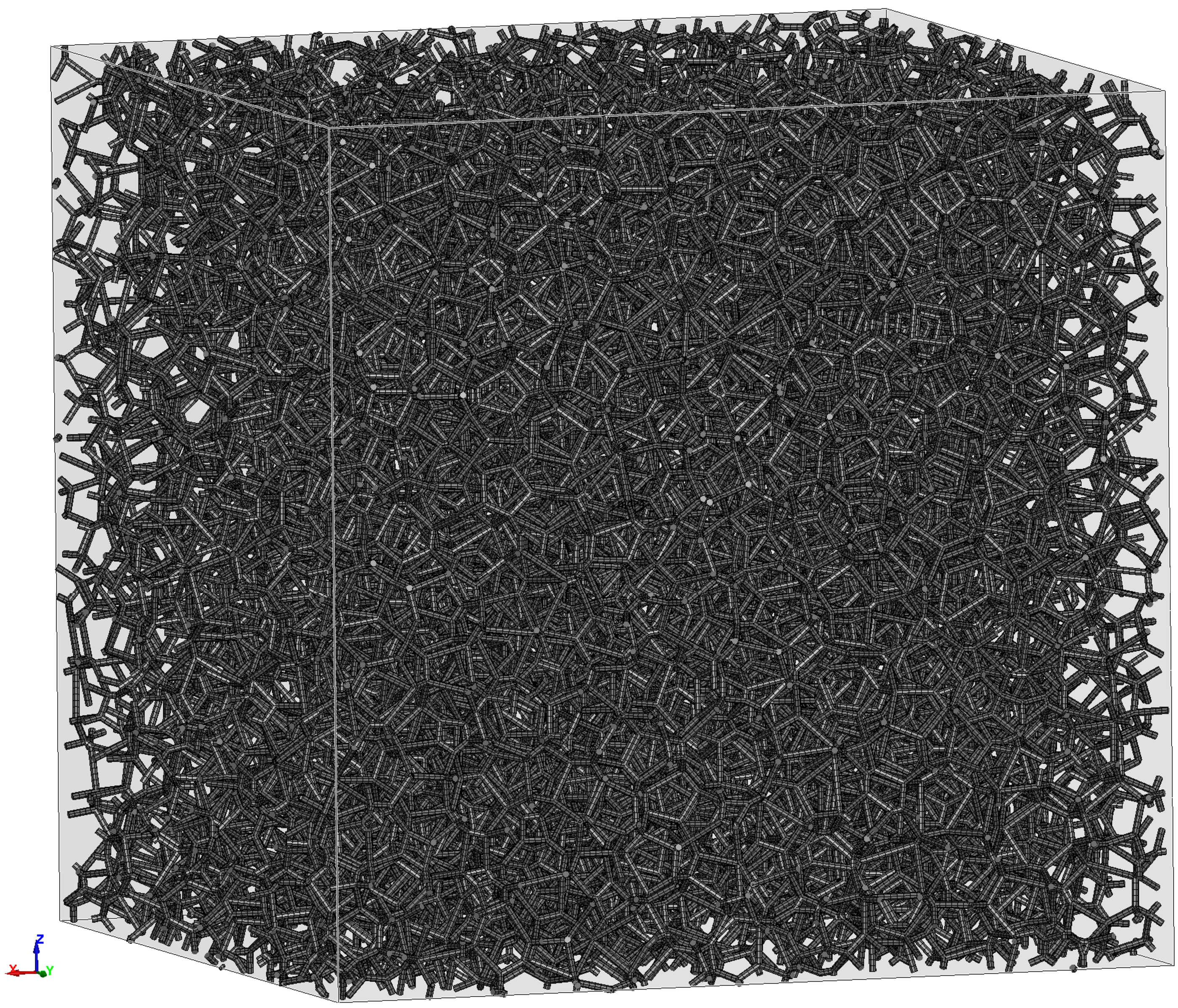}
  \caption{}
  \label{fig:RVE}
\end{subfigure}%
\begin{subfigure}[b]{0.48\textwidth}
  \centering
  \includegraphics[width=\linewidth]{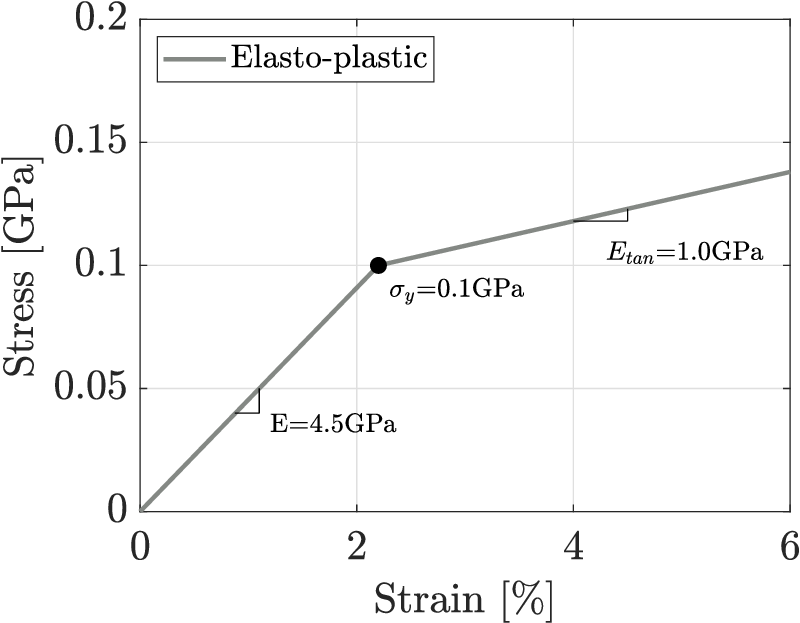}
  \caption{}
  \label{fig:elastoplastic}
\end{subfigure}
\caption{Illustration of (a) RVE with PBCs, and (b) Stress-strain diagram of elastoplastic material model.}
\label{fig:4}
\end{figure*}

The resulting Voronoi diagram is transformed to a cube-shaped representative volume element (RVE) with periodic boundary conditions (PBCs) as described in \cite{bib14}. An RVE in combination with PBCs can be used to obtain the homogenized macroscopic response of the material. The RVE of the microstructure model in Fig.~\ref{fig:modelflow}d with PBCs is illustrated in Fig.~\ref{fig:RVE}. The cell walls of the Voronoi network are discretized using Hughes-Liu beam elements with circular cross-section. The cross-section diameter obtained in the geometric modeling is assigned to the beam elements. The beam elements are meshed with an element size equal to its corresponding cross-section diameter in order to provide an element aspect ratio close to 1.0. For a detailed description on the generation of an RVE, refer \cite{bib13}. 

A bilinear elastoplastic material model, as illustrated in Fig.~\ref{fig:elastoplastic}, is used to characterize the behavior of cell walls with inuput parameters, namely Young's modulus ($E= 4.5$GPa from \cite{bib13}), yied stress ($\sigma_y=0.1$GPa) and tangent modulus ($\mathrm{E}_{\text{tan}}=1.0$GPa). The automatic general contact algorithm is used to model the beam-to-beam contact capturing interaction of cell walls. The model was subjected to monotonic and cyclic compression loading and the simulations were carried out using LS-DYNA implicit solver.


\section{Results}\label{sec3}
In this section, the effective (macroscopic) response of the microstructure model (RVE) under compression for different PSD, CWD and solid fraction is discussed.

\subsection{Macroscopic response}
\begin{figure*}
\centering
\begin{subfigure}[b]{0.6\textwidth}
  \centering
  \includegraphics[width=\linewidth]{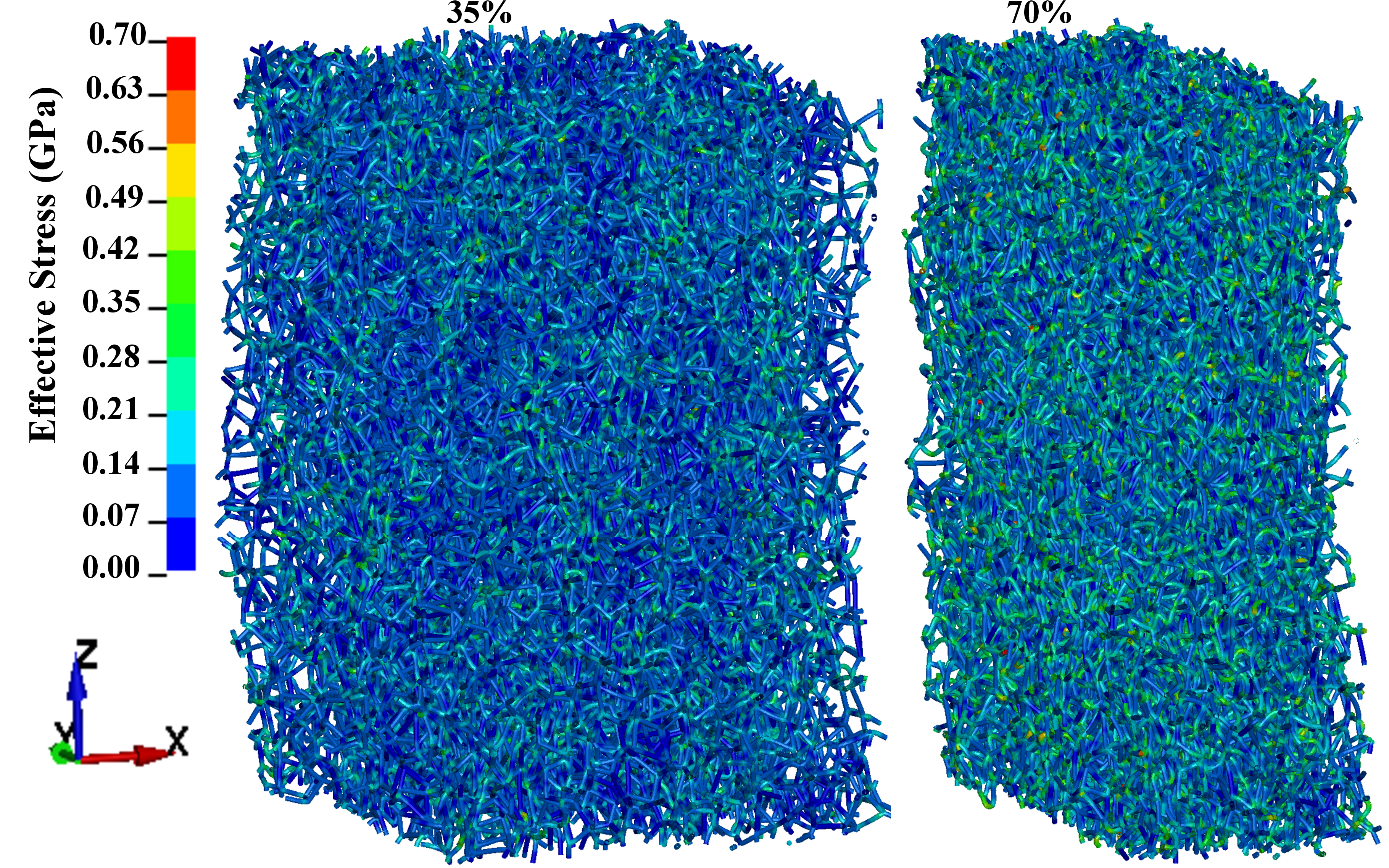}
  \caption{}
  \label{fig:RVE_deform}
\end{subfigure}%
\begin{subfigure}[b]{0.4\textwidth}
  \centering
  \includegraphics[width=\linewidth]{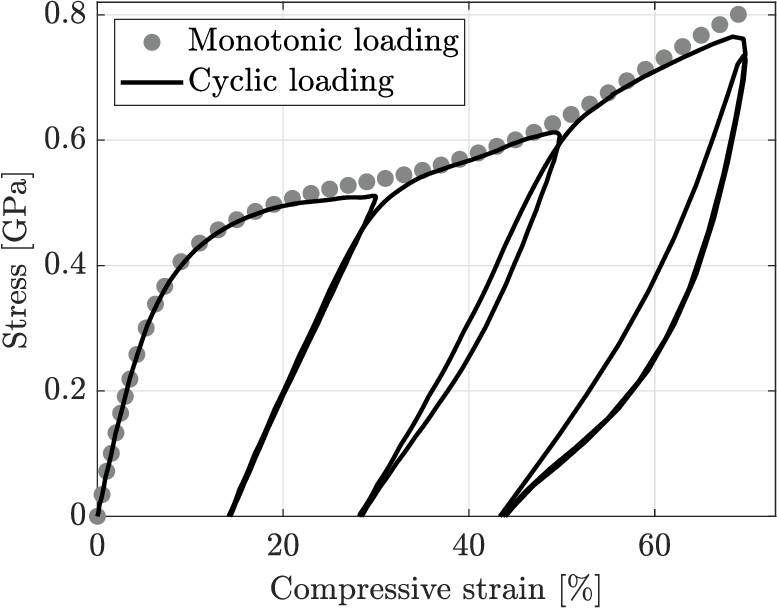}
  \caption{}
  \label{fig:RVE_response}
\end{subfigure}
\caption{(a) Von-Misses stress distribution under compression, and (b) Macroscopic stress-strain response under monotonic and cyclic compressive loading.}
\label{fig:4}
\end{figure*}

\begin{figure*}[h!]
\centering
\begin{subfigure}[b]{.4\textwidth}
  \centering
  \includegraphics[width=0.45\linewidth]{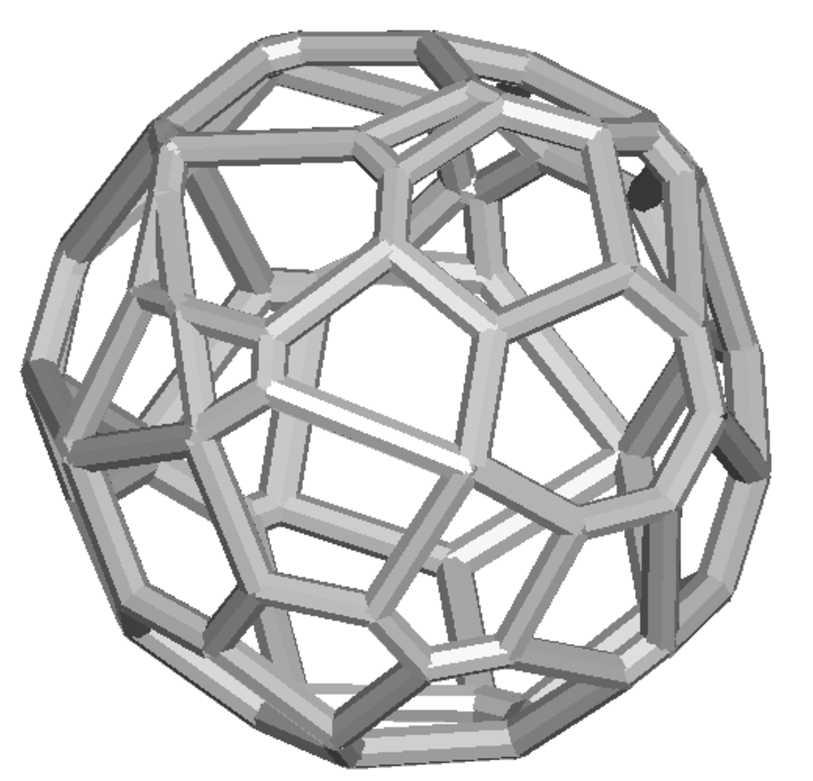}
  \caption{}
  \label{fig:poreUndeform}
\end{subfigure}%
\begin{subfigure}[b]{.3\textwidth}
  \centering
  \includegraphics[width=0.45\linewidth]{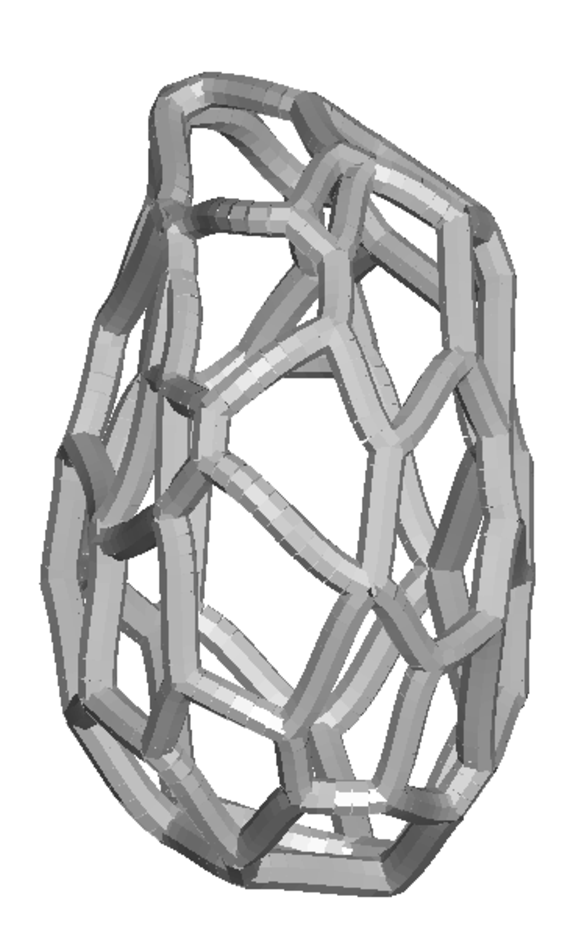}
  \caption{}
  \label{fig:poreDeform_1}
\end{subfigure}%
\begin{subfigure}[b]{.2\textwidth}
  \centering
  \includegraphics[width=0.45\linewidth]{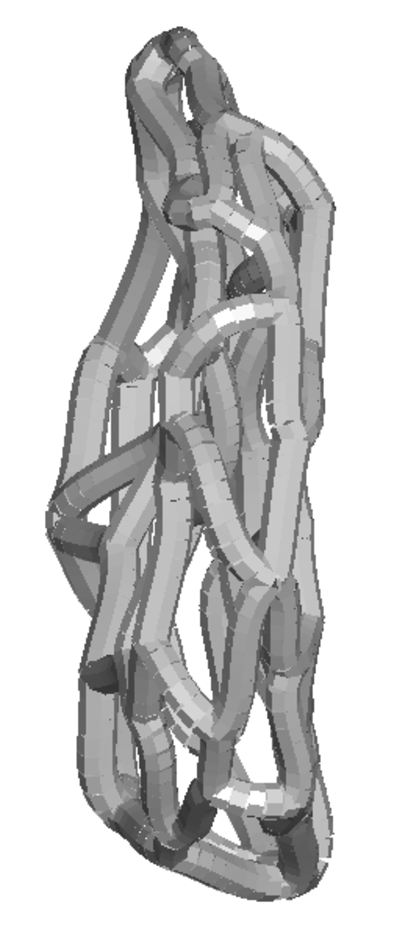}
  \caption{}
  \label{fig:poreDeform_2}
\end{subfigure}
\caption{Illustration of the cell wall behavior due to deformation of the structure under compression: (a) undeformed cell, (b) bending and buckling of cell walls, and (c) contact between cell walls after cell collapse.}
\label{fig:7}
\end{figure*}

\noindent
The deformation of the RVE structure (shown in Fig.~\ref{fig:modelflow}) and the corresponding contour plots representing the von-Misses effective stress distribution at 35\% and 70\% compressive strains are shown in Fig.~\ref{fig:RVE_deform}. From Fig.~\ref{fig:RVE_response}, it can be inferred that the RVE model captures the elastic as well as inelastic behavior, typically observed in such materials \cite{bib16}. Under compression, the energy is absorbed by the structure as the pore walls (Fig.~\ref{fig:poreUndeform}) experience buckling and bending (Fig.~\ref{fig:poreDeform_1}). Furthermore, once the cell walls buckle or elastically bend, the pores start to collapse resulting in a plateau in stress-strain diagram. Subsequently, the collapsed pores resist further deformation due to lack of available space referred to as densification (Fig.~\ref{fig:poreDeform_2}). 
\vspace{-1cm}

Under cyclic loading, the model captures the Mullins effect like behavior known from elastomers (i.e., the stress softening between the first and subsequent loadings) and a very small hysteresis (i.e., the stress softening between the unloading and the reloading of the subsequent cycle) along with the permanent set which is due to the irreversible damage in the microstructure.

\subsection{Representativeness of RVE}\label{sec:3.1.1}
In general, an RVE should have a size large enough such that any increase in its volume will be equally representative. Thus, minimizing the RVE size is desired typically to make the simulation computationally less expensive. 

\begin{figure*}[h!]
\centering
\begin{subfigure}[b]{.5\textwidth}
  \centering
  {\small \textbf{Symmetric beta PDF}}\par\medskip
  \includegraphics[width=0.9\linewidth]{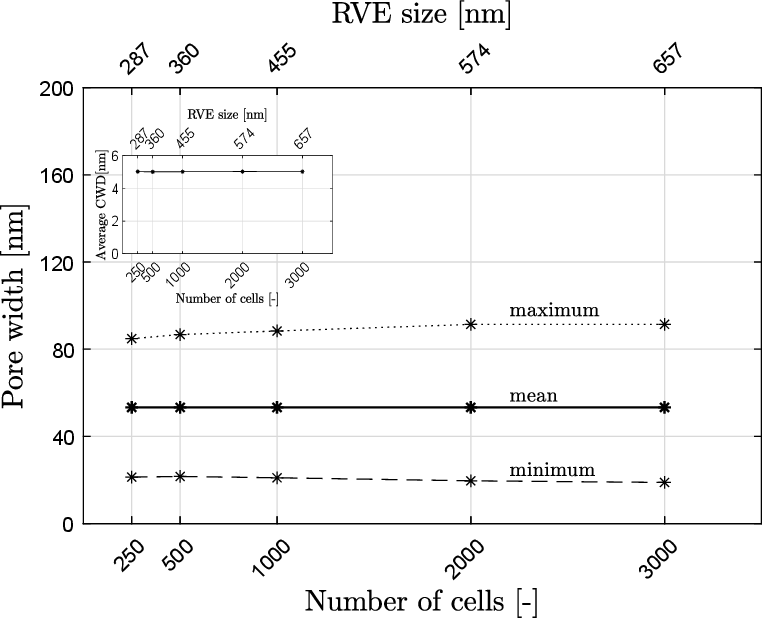}
  \caption{}
  \label{fig:varyingCells_pw_5_5}
\end{subfigure}%
\begin{subfigure}[b]{.5\textwidth}
  \centering
 {\small \textbf{Right skewed beta PDF}}\par\medskip
  \includegraphics[width=0.9\linewidth]{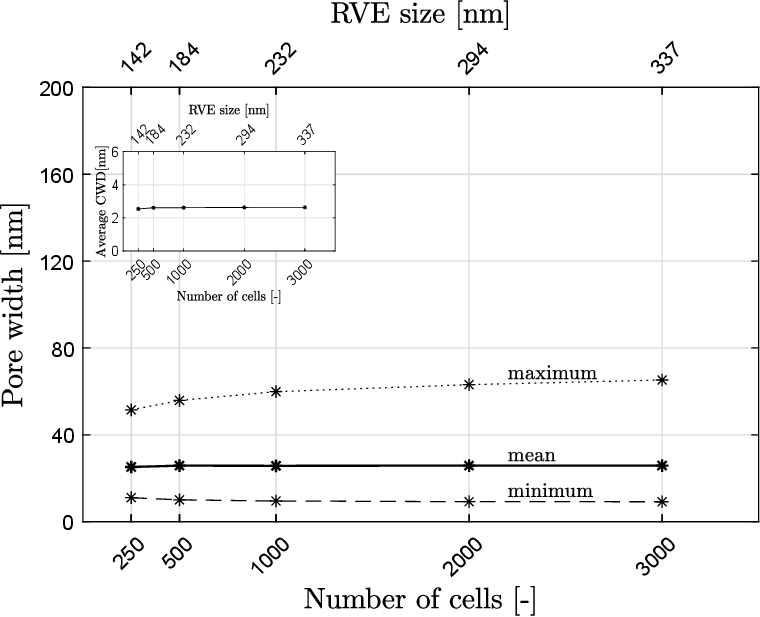}
  \caption{}
  \label{fig:varyingCells_pw_2_10}
\end{subfigure}
\vskip\baselineskip
\begin{subfigure}[b]{.5\textwidth}
  \centering
  \includegraphics[width=0.9\linewidth]{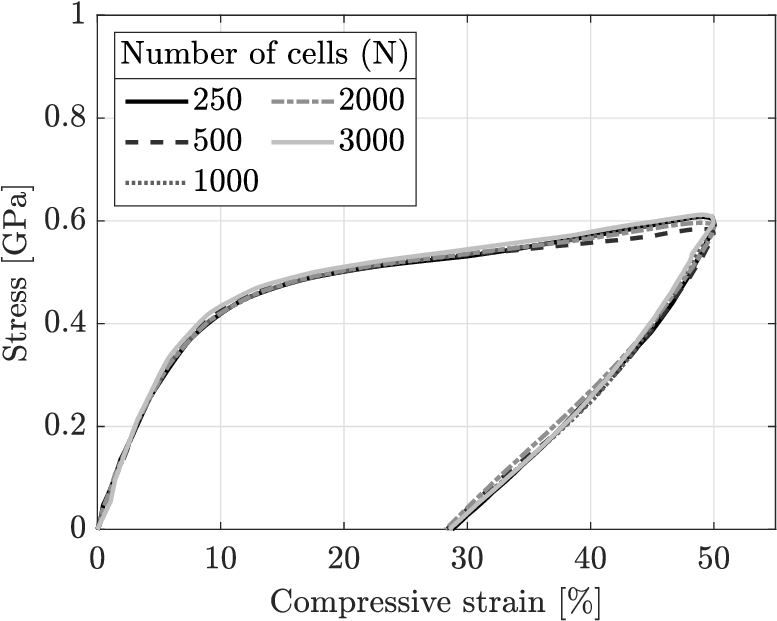}
  \caption{}
  \label{fig:varyingCells_ss_5_5}
\end{subfigure}%
\begin{subfigure}[b]{.5\textwidth}
  \centering
  \includegraphics[width=0.9\linewidth]{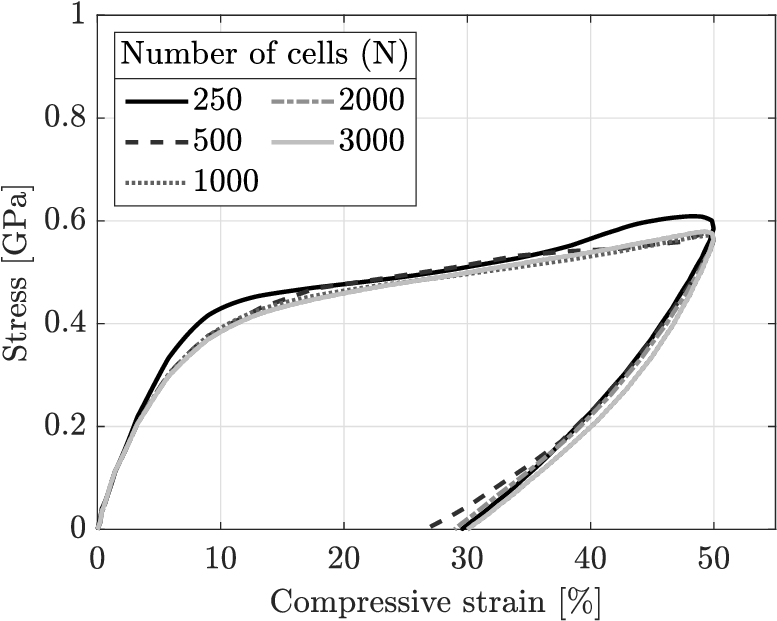}
  \caption{}
  \label{fig:varyingCells_ss_2_10}
\end{subfigure}
\caption{Illustration of evaluating the RVE size based on the convergence study of microstructural properties, (a) \& (b) maximum, minimum and mean pore-width with a subplot representing the average cell wall diameter, and  (d) \& (e) the effective macroscopic response, for symmetric beta PDF with $\alpha=5$, $\beta=5$ (on the left) and right skewed beta PDF with $\alpha=2$, $\beta=10$ (on the right).}
\label{fig:6}
\end{figure*}

\noindent
Therefore, the representativeness of an RVE is studied based on the microstructural properties and the effective mechanical properties resulting from the model. The statistical properties of the PSD as the mean and the pore-width range are compared for different RVE sizes and the respective structural parameter as CWD. According, the effective macroscopic properties are evaluated. Based on this study, the most favorable RVE size is chosen. Thus, RVEs with 250, 500, 1000, 2000 and 3000 cells, were generated based on a symmetric and a right skewed beta PDF and the constant solid fraction of 0.05. The corresponding statistical properties of the PSD of the structures are compared as shown in Fig.~\ref{fig:varyingCells_pw_5_5}~\&~\ref{fig:varyingCells_pw_2_10}. For the symmetric beta PDF (refer Fig.~\ref{fig:varyingCells_pw_5_5}), with the increasing number of cells from 250 to 3000,  the pore-width range changes from 21.32-84.82 nm to 18.86-91.42 nm, while resulting in a constant mean pore-width of 53.27$\pm0.02$ nm. For the right skewed beta PDF (refer Fig.~\ref{fig:varyingCells_pw_2_10}), the pore-width range changes from 10.62-45.69 nm to 9.2-65.26 nm in this case. However, the calculated average CWD is the same for the given PDF irrespective of the RVE size (refer to inner plot in Fig.~\ref{fig:varyingCells_pw_5_5} \&~\ref{fig:varyingCells_pw_2_10}). Even though there is not much variation in the minimum pore-width while the maximum pore-width increases by 20\% resulting in the change of the elastic and the inelastic response (refer Fig.~\ref{fig:varyingCells_ss_2_10}). From Fig.~\ref{fig:varyingCells_ss_5_5}~\&~\ref{fig:varyingCells_ss_2_10}, it becomes apparent that the effective macroscopic response is identical for converged statistical properties. Therefore, for the same solid fraction and mean pore-size, as well as the same CWD, the minimum representative size of the RVE is identified.
   
\subsection{Influence of cell wall diameter (CWD)}\label{sec:3.1.2}
\begin{figure*}[h!]
\centering
\begin{subfigure}[b]{.5\textwidth}
  \centering
  \includegraphics[width=0.9\linewidth]{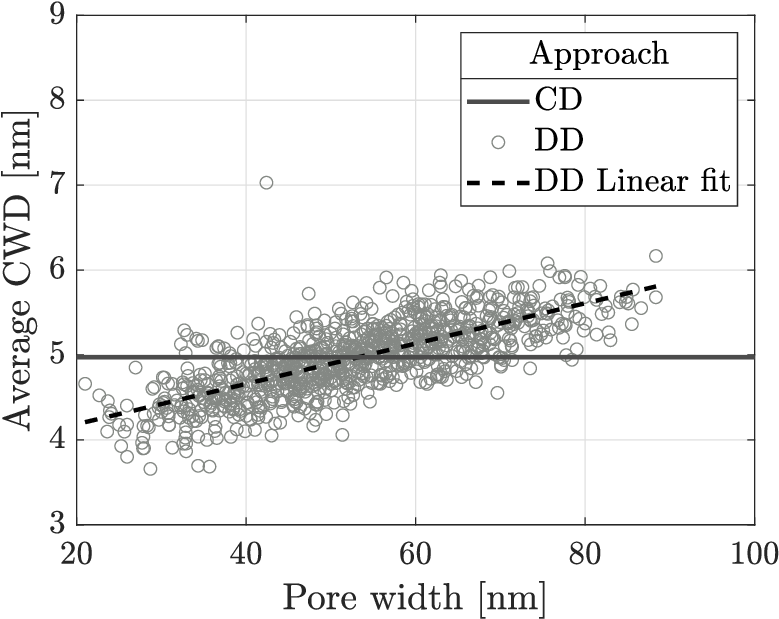}
  \caption{}
  \label{fig:cellwalldiaVsporesize}
\end{subfigure}%
\begin{subfigure}[b]{.5\textwidth}
  \centering
  \includegraphics[width=0.9\linewidth]{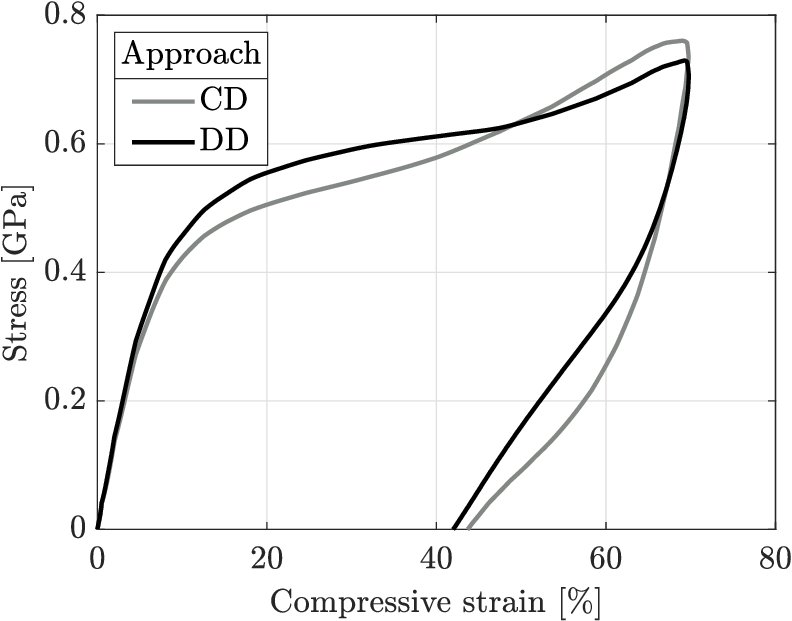}
  \caption{}
  \label{fig:sim_cellwalldiaVsporesize}
\end{subfigure}
\caption{(a) Average cell wall diameter for each pore-width, and (b) bulk response of the structure with CWD determined using CD and DD approach}
\label{fig:7}
\end{figure*}

The average CWD determined using CD and DD approach (as explained in Sect.~\ref{sec:2.3.1}) for different pore-sizes in the structure generated for the given symmetric beta PDF and solid fraction is illustrated in Fig.~\ref{fig:cellwalldiaVsporesize}.

In DD approach, the resulting CWD increases with the pore-size whereas a constant CWD is defined to the entire structure in the CD approach. The bulk response of the structure with CWD determined using both the approaches differs under compression loading and unloading, as shown in Fig.~\ref{fig:sim_cellwalldiaVsporesize}. In Fig.~\ref{fig:cellwalldiaVsporesize}, we notice that the larger pore-widths have smaller CWD in CD approach which results in a stronger stress softening in comparison to DD approach (see Fig.~\ref{fig:sim_cellwalldiaVsporesize}), as larger cells begin to deform first under small strains. Similarly, sooner densification is obtained with CD approach in comparison with DD approach. This is because the smaller cells have larger CWD in CD approach resulting in early contact of cell walls after pore collapse. Accordingly, the densification is influenced by the smaller cells. Although the porosity of the structure is same, the method of the CWD definition influences the bulk response of the structure.

\begin{figure*}[h!]
\centering
\begin{subfigure}[b]{.5\textwidth}
  \centering
  \includegraphics[width=\linewidth]{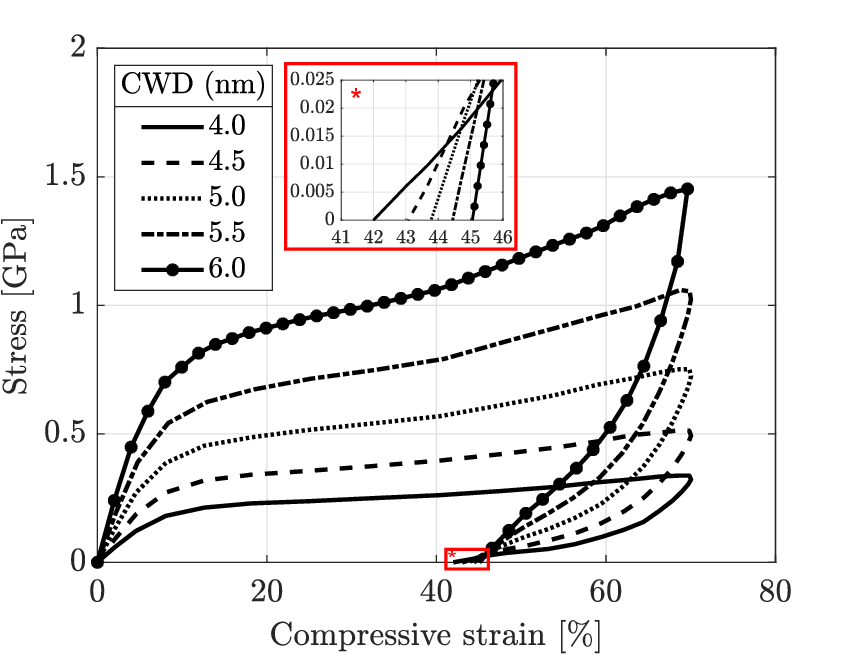}
  \caption{}
  \label{fig:sim_cellwalldia}
\end{subfigure}%
\begin{subfigure}[b]{.5\textwidth}
  \centering
  \includegraphics[width=0.925\linewidth]{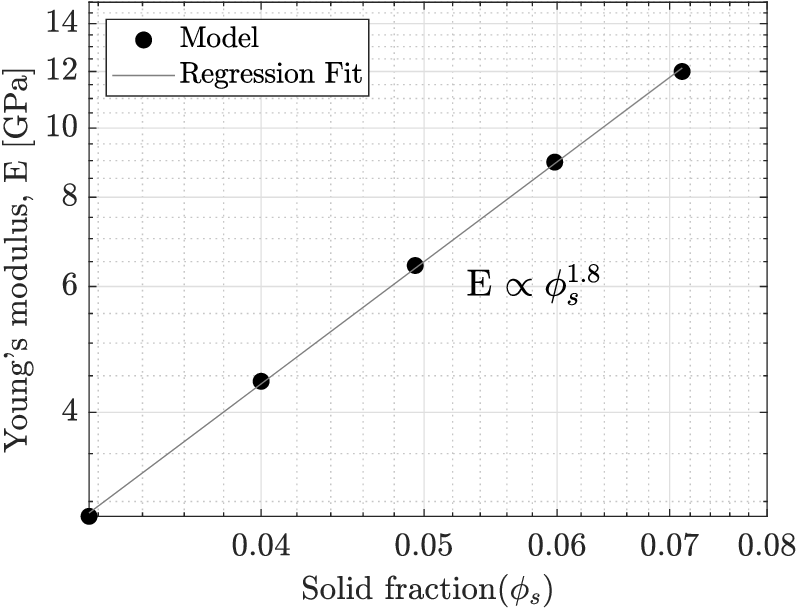}
  \caption{}
  \label{fig:cellwalldia}
\end{subfigure}
\caption{(a) Bulk response and (b) bulk stiffness of the structure with increasing constant CWD (resulting in increasing solid fraction).   }
\label{fig:7}
\end{figure*}

The influences of constant CWD on the bulk constitutive response of the structure is shown in Fig.~\ref{fig:sim_cellwalldia}. It can be inferred that the solid fraction increases with the CWD and thereby decreases the porosity of the structure, resulting in a stiffer mechanical response.
Indeed, with the increasing CWD the bulk elastic stiffness (E) and the specific energy density increase, while the span of plateau regime and the strain at which the densification begins decrease. This effects are in line with previous constitutive modeling results \cite{bib18}. On the other hand, a very small variation in the residual deformation is observed with increasing CWD. It is also inferred that the relationship between Young's modulus and the solid fraction ($\phi_s$) can be characterized by a scaling law with a scaling exponent of 1.8 (refer Fig.~\ref{fig:cellwalldia}), which is close to 2 as given by open porous foam models \cite{bib9}.

\subsection{Influence of pore-size distribution (PSD)}\label{sec:3.2}
The pore-size distribution varies for different PDFs with different statistical properties (mean and standard deviation) and shape. It is essential to study the influence of the PSD on other geometric parameters, namely solid fraction and CWD, and the bulk mechanical response of the structure.
\subsection*{Different mean pore-width and same shape of PSD:}
\begin{figure*}[h!]
\centering
\begin{subfigure}[b]{.5\textwidth}
  \centering
  \includegraphics[width=\linewidth]{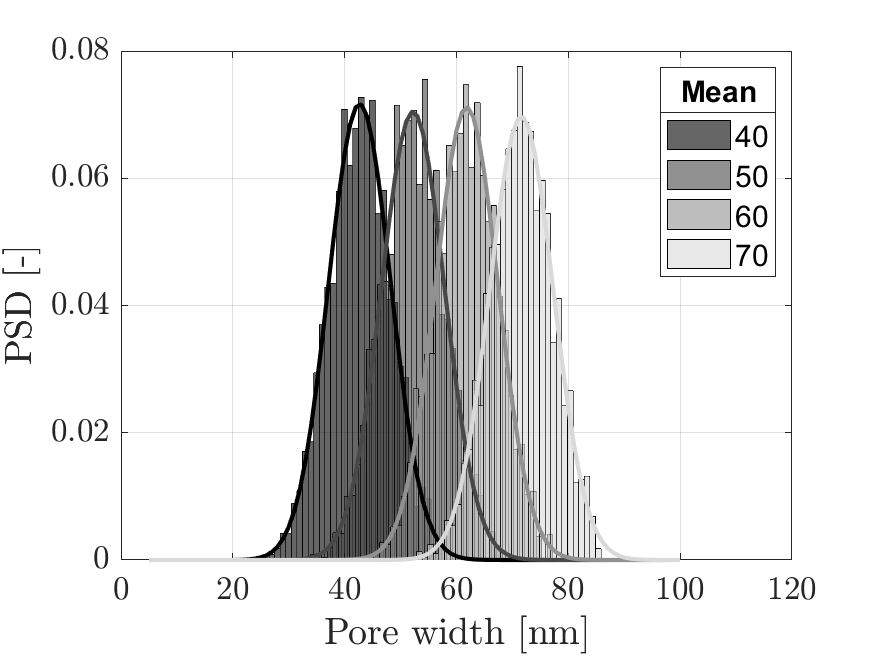}
  \caption{}
  \label{fig:varyingmean_psd}
\end{subfigure}%
\begin{subfigure}[b]{.5\textwidth}
  \centering
  \includegraphics[width=\linewidth]{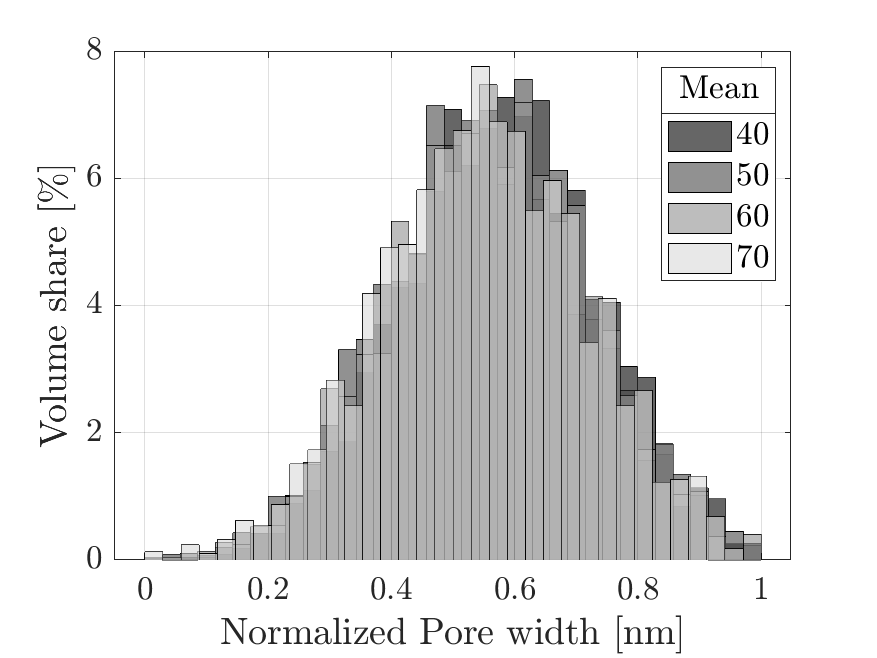}
  \caption{}
  \label{fig:varyingmean_psd_norm}
\end{subfigure}%
\vskip\baselineskip
\begin{subfigure}[b]{.5\textwidth}
  \centering
  \includegraphics[width=0.9\linewidth]{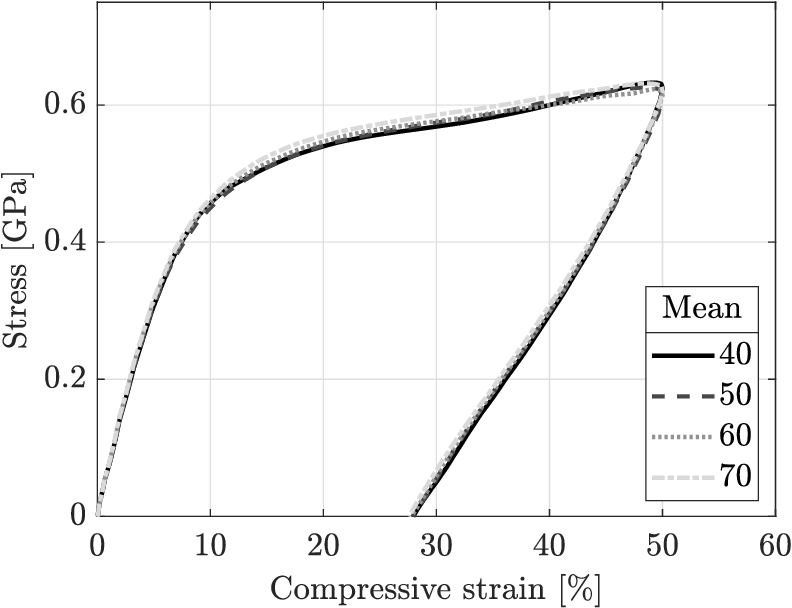}
  \caption{}
  \label{fig:varyingmean_sim}
\end{subfigure}
\caption{Comparison of (a) PSDs based on normal PDFs, (b) PSDs with pore-widths in normalized scale, and (c) bulk response of the structure with PSDs corresponding to Fig.~\ref{fig:varyingmean_psd}, for different mean pore-widths and constant standard deviation.}
\label{fig:7}
\end{figure*}

For any distribution, the shape of PSD remains unchanged for a constant standard deviation. PSDs with the constant standard deviation and different mean pore-widths (namely 40, 50, 60, and 70 nm) based on the normal PDF are compared in Fig.~\ref{fig:varyingmean_psd}. Due to the symmetric property of the normal PDF, the PSDs are also symmetric and have similar shape, but pore-width range (minimum and maximum) is shifted according to the given mean pore-width. As a consequence, the resulting CWD (for a given solid fraction of 0.05) and RVE size also increase to the equivalent percentage of increase of the mean pore-width (refer Table~\ref{tab:Norm_parameters}).

\begin{table}[h!]
\begin{center}
\begin{minipage}{\textwidth}
\caption{Geometric properties of the structures with PSDs corresponding to Fig.~\ref{fig:varyingmean_psd}, with constant standard deviation and number of cells, N=1000}\label{tab:Norm_parameters}%
\begin{tabular*}{\textwidth}{@{\extracolsep{\fill}}ccccc@{\extracolsep{\fill}}}
\toprule
\multicolumn{3}{c}{Pore Width}&Constant CWD$^a$&RVE size \\ \cmidrule{1-3}
Mean (input) & Minimum & Maximum  &   &  \\
\midrule
40  & 22 & 58 & 3.615 & 332  \\
50  & 32 & 68 & 4.462 & 410  \\
60  & 42 & 78 & 5.315 & 490  \\
70  & 52 & 88 & 6.175 & 570  \\
\midrule
\end{tabular*}
\footnotetext{Note: The unit of all the parameters is nm}
\footnotetext{$^a$ A constant CWD is determined for all structures with a constant solid fraction of 0.05.}
\end{minipage}
\end{center}
\end{table}

The resulting bulk responses of the structures are similar under loading and unloading, as shown in Fig.~\ref{fig:varyingmean_sim}. This is due to the fact that all the structures have the identical PSD in the normalized scale as shown in Fig.~\ref{fig:varyingmean_psd_norm}, resulting in the same CWD and RVE size. Therefore, geometric properties are scaled with the pore-width and the corresponding structure will show similar bulk response for a given shape of PSD and solid fraction. This extends the results by Aney and Rege \cite{bib12} (where such behavior was shown for only linear-elastic regime) to large deformation.

\subsection*{Different shape of PSD and same mean pore-width:}
\begin{figure*}[ht!]
\centering
\begin{subfigure}[b]{0.5\textwidth}
  \centering
  \includegraphics[width=0.925\linewidth]{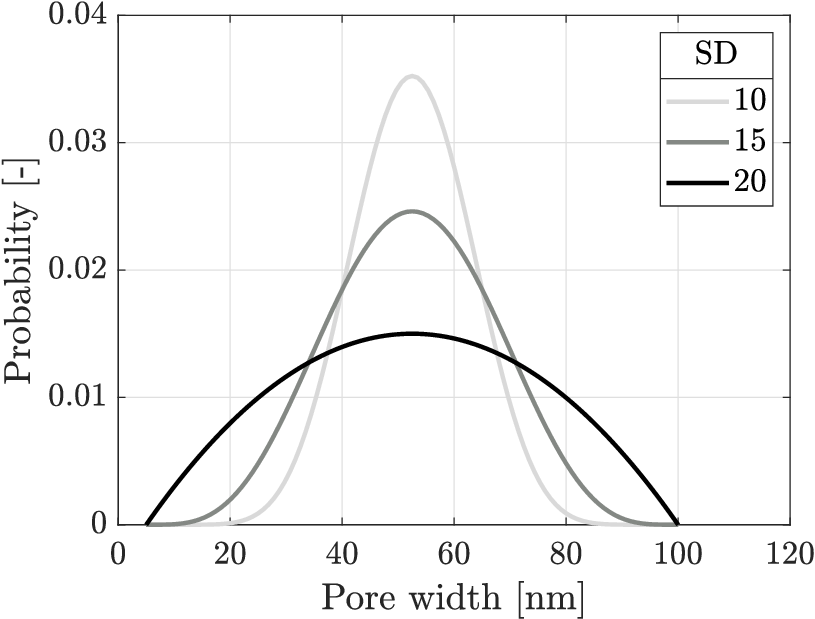}
  \caption{}
  \label{fig:varyingSD_pdf}
\end{subfigure}%
\begin{subfigure}[b]{0.5\textwidth}
  \centering
  \includegraphics[width=\linewidth]{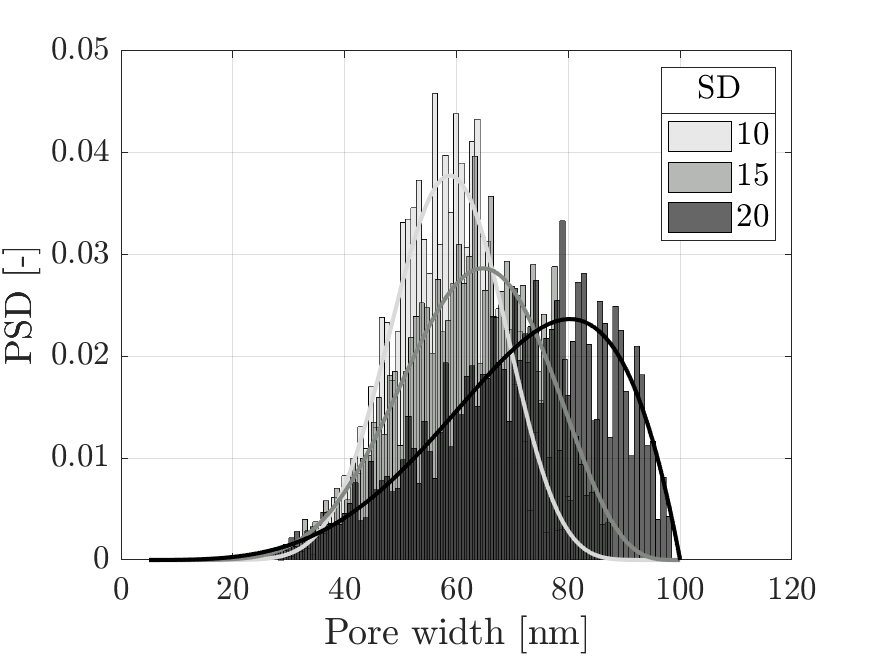}
  \caption{}
  \label{fig:varyingSD_psd}
\end{subfigure}%
\vskip\baselineskip
\begin{subfigure}[b]{0.5\textwidth}
  \centering
  \includegraphics[width=0.9\linewidth]{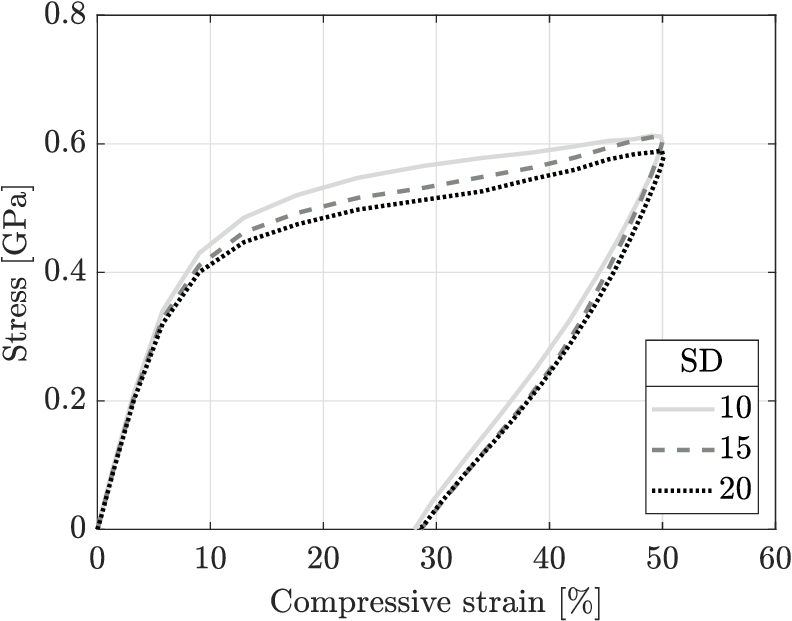}
  \caption{}
  \label{fig:Sim_varyingSD}
\end{subfigure}
\caption{Comparison of (a) beta PDFs, (b) PSDs corresponding to the PDFs, and (c) bulk response of the structures, for different standard deviation and constant mean pore-width.}
\label{fig:11}
\end{figure*}

Different PSDs with constant mean pore-width and different standard deviation (SD) are compared in Fig.~\ref{fig:11} using symmetric beta PDF. The latter PDFs with $\alpha = \beta = $10, 5 and 2 with SDs = 10, 15 and 20 respectively and constant mean pore-width = 53 nm are shown in Fig.~\ref{fig:varyingSD_pdf}. The PSD of the Voronoi structures is shown in Fig.~\ref{fig:varyingSD_psd} in comparison with the desired PSDs corresponding to the PDFs in Fig.~\ref{fig:varyingSD_pdf}. 

By increasing the standard deviation, the maximum pore-width increases, while the minimum pore-width decreases (see Table~\ref{tab:Beta_parameters}), resulting in different shape of the PSDs. As a consequence, we see a marginal difference in the bulk response of the structures above the elastic regime, see Fig.~\ref{fig:Sim_varyingSD}. The bulk stiffness of all three structures is the same in the elastic regime, as they have the same solid fraction. However, above the elastic regime, we notice a softening response as the standard deviation of the PDF increases, even though the CWD also increases. This is because the shape of the PSD changes significantly in such a way that the contribution of the larger cells dominates as the standard deviation of the PDF increases. Therefore, cells of larger size begin to undergo bending sooner, resulting in the softening response. However, there is no influence on the specific energy density and residual deformation under unloading. 

\begin{table}[h!]
\begin{center}
\begin{minipage}{\textwidth}
\caption{Geometric properties of the structures with PSDs corresponding to Fig.~\ref{fig:11}, with constant mean pore width and number of cells, N=1000}\label{tab:Beta_parameters}%
\begin{tabular*}{\textwidth}{@{\extracolsep{\fill}}ccccc@{\extracolsep{\fill}}}
\toprule
Standard deviation & \multicolumn{2}{c}{Pore Width} & Constant CWD$^a$& RVE size \\  \cmidrule{2-3} 
SD (input) & Minimum & Maximum  &   &  \\
\midrule
10  & 26   & 81 & 4.82 & 440  \\
15  & 21 & 88 & 5.03 & 455  \\
20  & 17 & 98 & 5.49 & 490  \\
\midrule
\end{tabular*}
\footnotetext{Note: The unit of all the parameters is nm}
\footnotetext{$^a$ A constant CWD is determined for all structures with a constant solid fraction of 0.05.}
\end{minipage}
\end{center}
\end{table}


\subsection*{Different PSD and same CWD:}
In order to study only the influence of the PSD neglecting the effect of the CWD on the bulk response, different PSDs using normal and beta PDFs (referred to as PSD- 1,2 \& 3) resulting in a structure with same average CWD and solid fraction are compared in this section.

\begin{figure*}[ht!]
\centering
\begin{subfigure}[b]{0.5\textwidth}
  \centering
  \includegraphics[width=1.08\linewidth]{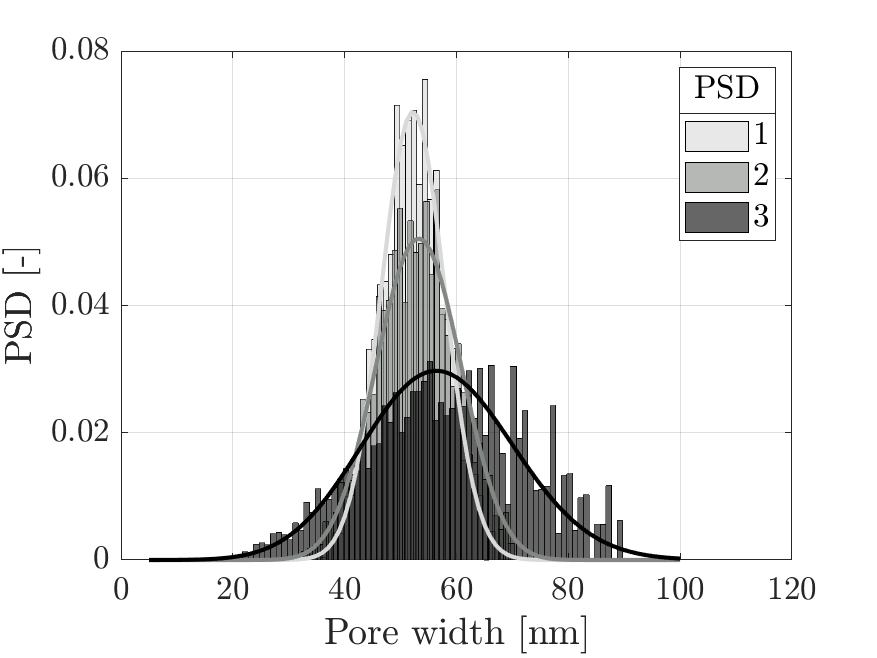}
  \caption{}
  \label{fig:varyingPSD}
\end{subfigure}%
\begin{subfigure}[b]{0.5\textwidth}
  \centering
  \includegraphics[width=\linewidth]{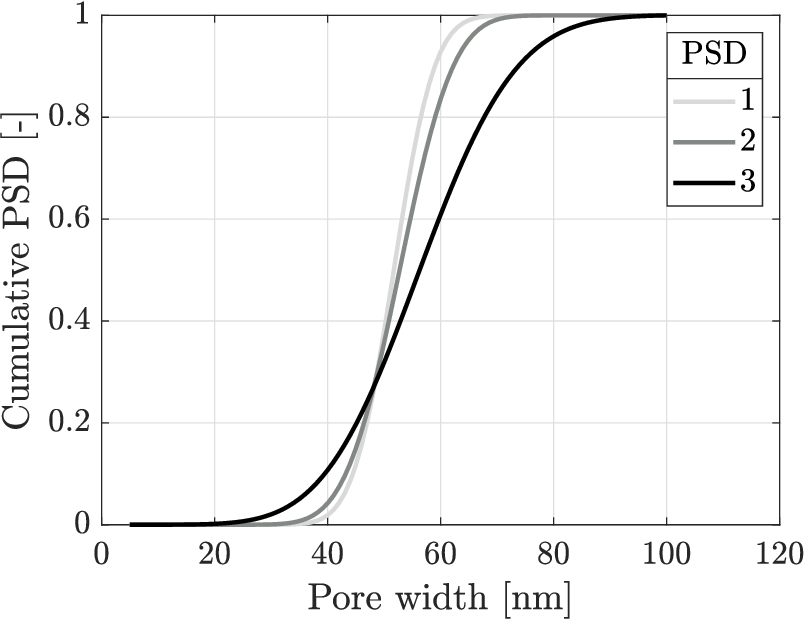}
  \caption{}
  \label{fig:varyingPSD_cum}
\end{subfigure}
\vskip\baselineskip
\begin{subfigure}[b]{0.5\textwidth}
  \centering
  \includegraphics[width=\linewidth]{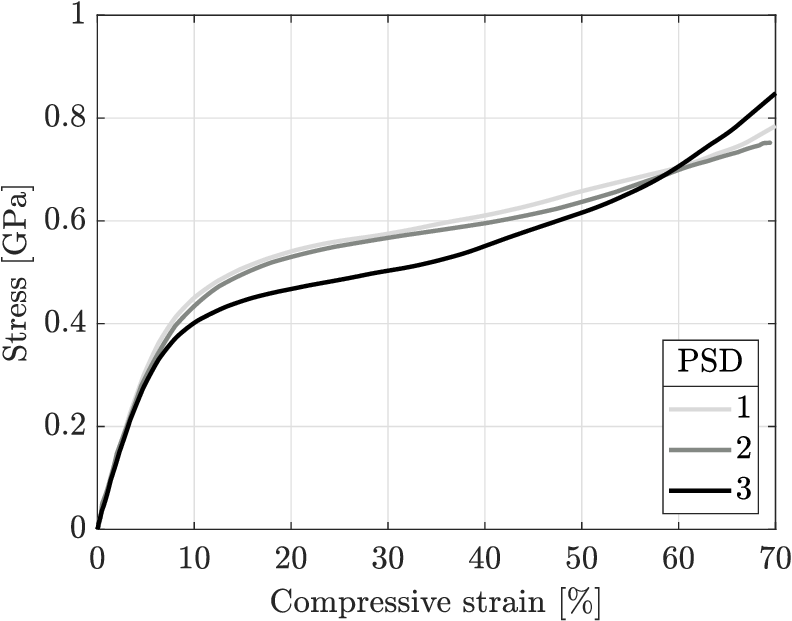}
  \caption{}
  \label{fig:Sim_varyingPSD}
\end{subfigure}%
\begin{subfigure}[b]{0.5\textwidth}
  \centering
  \includegraphics[width=\linewidth]{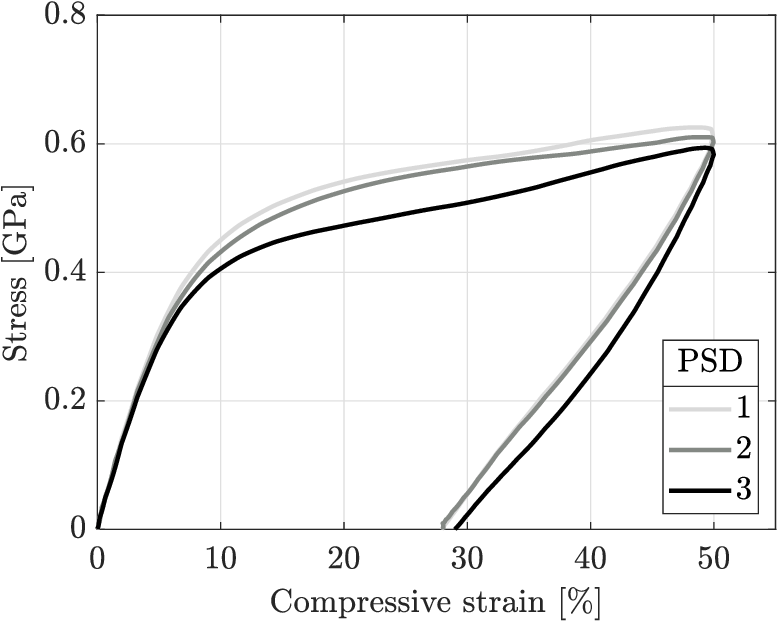}
  \caption{}
  \label{fig:Sim_varyingPSD_cyclic}
\end{subfigure}
\caption{Comparison of (a) PSDs and (b) cumulative PSDs with different standard deviations and the same mean, and the bulk response of the structures with PSD in Fig.~\ref{fig:varyingPSD} under compression with (c) monotonic loading up to 70\% strain and (d) loading up to and unloading at 50\% strain.}
\label{fig:15}
\end{figure*}

\begin{table}[h!]
\begin{center}
\begin{minipage}{\textwidth}
\caption{Geometric properties of the structures with PSDs corresponding to Fig.~\ref{fig:varyingPSD}, with the same average CWD of 4.5nm and solid fraction of 0.05 with 1000 number of cells.}\label{tab:Varying_PSD}%
\begin{tabular*}{\textwidth}{@{\extracolsep{\fill}}ccccccc@{\extracolsep{\fill}}}
\toprule
\multicolumn{2}{c}{PSD} & \multicolumn{3}{c}{Pore width [nm] } & SD &  RVE size\\\cmidrule{1-5}
No. &  Input & Mean  & Minimum & Maximum  &  & [nm]   \\
\midrule
1  & Normal  & 50 & 33 & 68 & 5.8 & 410 \\
   & ($\mu$=50,~$s$=7)& & & & & \\
2  & Beta  & 50  & 28 & 70 & 7.3  & 410   \\
   & ($\alpha$=17.6,~$\beta$=20)& & & & &\\

3  & Normal   &  45 & 14 & 90 & 15& 390 \\
   & ($\mu$=40,~$s$=21)& & & & & \\
\midrule
\end{tabular*}
\end{minipage}
\end{center}
\end{table}

The stastical parameters of three different PDFs and corresponding geometric properties of the resulting structures are shown in Table~\ref{tab:Varying_PSD}, whose PSDs and the corresponding cumulative PSDs are shown in Fig.~\ref{fig:varyingPSD} \& \ref{fig:varyingPSD_cum}. The RVE size of all the structures is 400$\pm$10 nm. The structures have different pore-width range, mean pore-width and SD with an average CWD of 4.5nm and solid fraction of 0.05. From Fig.~\ref{fig:Sim_varyingPSD}, we see a significant difference in the bulk response of the structures. Further, we compared the PDFs and the corresponding statistical properties of the structures to study its respective bulk mechanical response. Accordingly, the structure with smaller mean pore-width shows less stiffer response, which is contradictory to the study in \cite{bib12}. However, from the cumulative PSD in Fig.~\ref{fig:varyingPSD_cum}, we observe that the volume contribution of larger cells in PSD-3 is stronger in comparison to PSD-1~\& -2, despite of smaller mean pore-width in PSD-3. Therefore, PSD-3 shows softening response above 10\% strain due to cell wall deformation of the larger cells. Similarly, we see a higher volume contribtution of smaller cells below the mean pore-width in PSD-3 resulting in an earlier densification compared to PSD- 1\& -2.  Although PSD-1 \& -2, have same mean, the shape of the PSD and cumulative PSD differs due to different standard deviation significantly influencing the bulk response at strain above 10\%. However, it appears to have a very slight influence on the specific energy density and permanent set under unloading. 

\section{Discussion}\label{sec4}
The computational model representing the realistic pore morphology of porous materials and capturing the important structural properties (PSD, solid fraction and pore wall thickness) is generated using RCPPS and LVT \cite{bib13}. The pore morphology of the computational model is described by the PSD which can be derived from the given PDF (Sect.~\ref{sec:2.2}). The CWD of the structure can be calculated for a given solid fraction. Accordingly, RVEs of different sizes inheriting the complete range of PSD show similar bulk responses. The minimal RVE size is identified by a convergence study on the effective geometric and mechanical properties of the structure.  The computational model captures both the elastic and inelastic phenomena of a typical porous material. The solid fraction also has a major influence on the specific energy density of the structure and permanent set. Indeed, the solid fraction increases with CWD which in turn raises the elastic modulus while reducing the plateau range and the compressive strain at which the densification takes place. The solid fraction represents a dimensionless quantity and is convenient to assign as a model parameter, based on which the CWD can be calculated for a given RVE size by reverse engineering as presented in \cite{bib13}. In addition, the method of defining the CWD (using CD and DD approach) also influences the bulk response. Therefore, further experimental investigation is needed to study the actual distribution of CWD of real porous structures. For a constant solid fraction and PSD with same shape, the structure shows similar bulk response at different scales. This is because the average CWD and RVE sizes scales according to the given pore-width interval and solid fraction. This could be better explained with an example of idealized square shaped cell, as first proposed in \cite{bib17}.

\begin{table}[ht!]
\begin{center}
\begin{minipage}{\textwidth}
\caption{Geometric properties of square shaped cell with cell wall length (l) and the same solid fraction.}\label{tab:square_props}%
\begin{center}
\begin{tabular*}{0.75\textwidth}{@{\extracolsep{\fill}}ccc@{\extracolsep{\fill}}}
\toprule
Cell wall length (l)  &  Cell width ($\sqrt{2}l$) & CWD ($d_f$)\\

[nm]  &  [nm]  & [nm]   \\
\midrule
5  & 7.07  & 0.73 \\
10  & 14.14  & 1.46   \\
20  & 28.28   &  2.92\\
\midrule
\end{tabular*}
\end{center}
\end{minipage}
\end{center}
\end{table}

\begin{figure}[h!]
\centering
\begin{subfigure}[b]{.4\textwidth}
  \centering
  \includegraphics[width=0.5\linewidth]{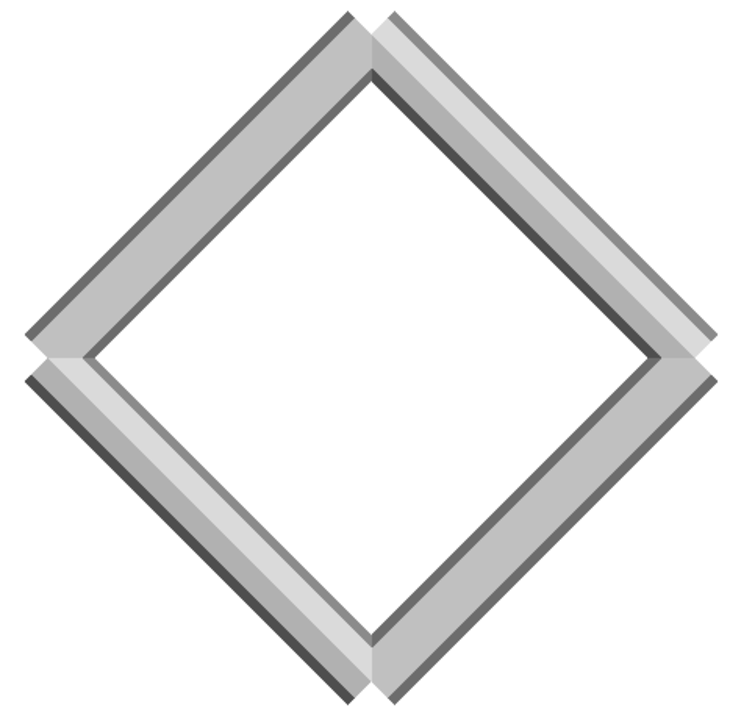}
  \caption{}
  \label{fig:squareUndeform}
\end{subfigure}%
\begin{subfigure}[b]{.3\textwidth}
  \centering
  \includegraphics[width=0.4\linewidth]{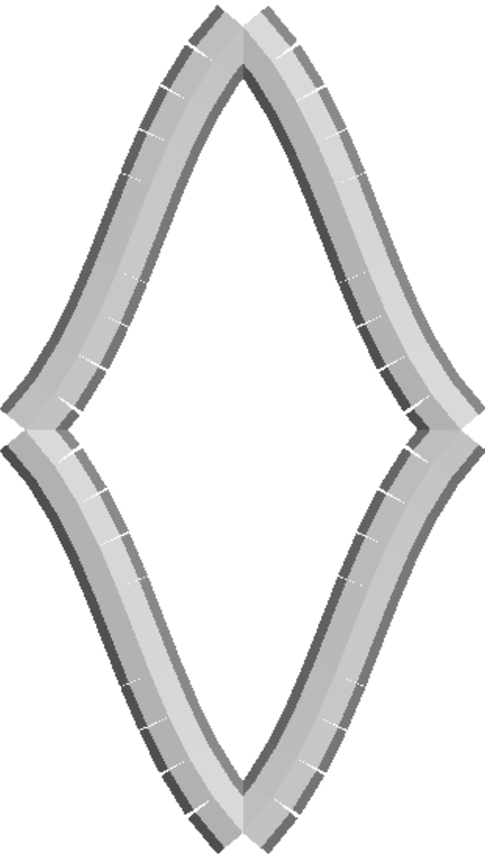}
  \caption{}
  \label{fig:squareeDeform_1}
\end{subfigure}%
\begin{subfigure}[b]{.2\textwidth}
  \centering
  \includegraphics[width=0.3\linewidth]{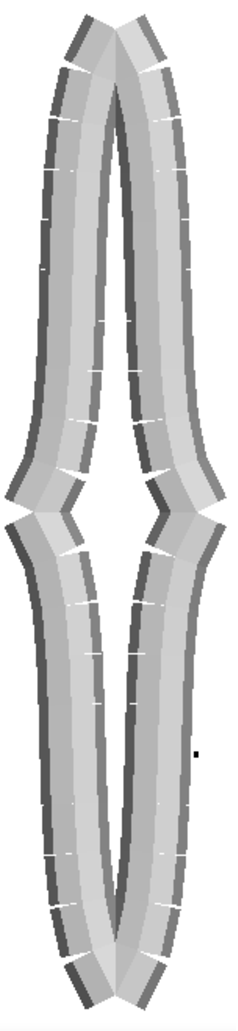}
  \caption{}
  \label{fig:squareDeform_2}
\end{subfigure}
\caption{Illustration of the cell wall behavior of a square shaped cell due to tension along the diagonal of the square: (a) undeformed cell, (b) deformation at 25\% strain, and (c) deformation at 50\% strain.}
\label{fig:square_deform}
\end{figure}

\begin{figure}[h!]
    \centering 
    \includegraphics[width=0.6\linewidth]{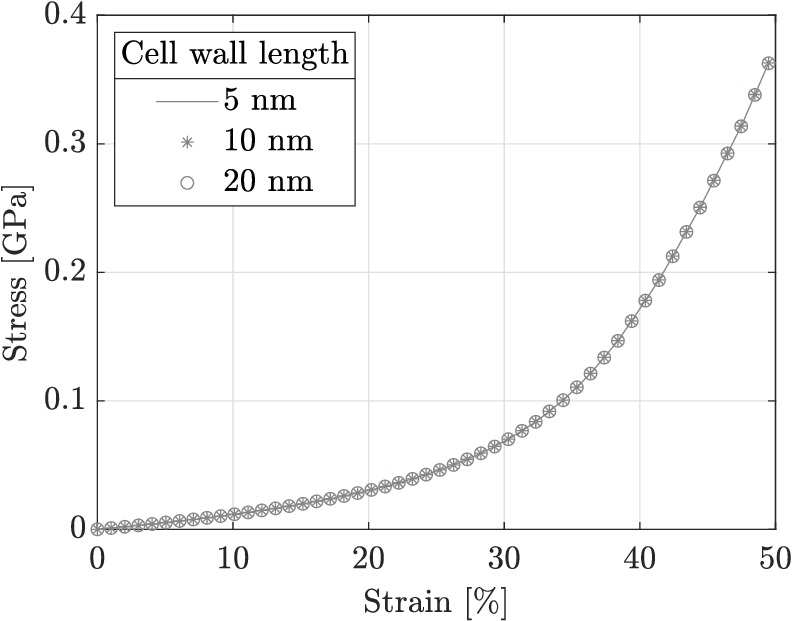}
    \caption{Stress-strain response resulting from a FE model of a square shaped cell subjected to tension along the diagonal. }
    \label{fig:square_response}
\end{figure}

 Let us consider a square shape cell with different side lengths $l=$ 5, 10 and 20 nm but with same solid fraction. The corresponding diameter ($d_f$) of the cell walls with circular cross-section for different cell sizes is tabulated in Table~\ref{tab:square_props}.
Note that for a given solid fraction, the CWD is scaled twice if the cell width (diagonal of the square cell) is scaled twice. The normal stress $\sigma$ in a square shape cell subjected to tension along its diagonal can be calculated as  \cite{bib15}

\begin{equation}
\label{eq:normalstress}
\sigma(\lambda,l) = E(\lambda-1) \sin\varphi \left[ \sin\varphi + \frac{3 \cos\varphi d_f}{2l}\right],
\end{equation}

where $\lambda$ is the stretch along the diagonal axis, $l$ is the side length, $\varphi$ is the angle of a single cell wall with respect to horizontal axis, $d_f$ is the cross-sectional diameter of the cell wall. As the $d_f$ and $l$ are scaled up with the same factor (see Table~\ref{tab:square_props}), the fraction $d_f/l$ in eq.~\ref{eq:normalstress} leads to same normal stress for square shaped cell of different sizes but the same solid fraction and pore shape.  In order to see the response of the square shaped cell at high strains, a FE simulation is carried out with a linear elastic material. Fig.~\ref{fig:square_deform} shows deformed configuration of a square shaped cell subjected to tension in the diagonal direction. From Fig.~\ref{fig:square_response}, we infer that even at higher strains, the structure shows similar response. 
  
On the other hand, the structure with different PSD but with same average CWD and solid fraction influences the bulk response significantly. According to the cumulative PSDs and respective bulk response, the elastic and the plateau regime in the stress-strain response of the porous materials are influenced by the larger cells, while the densification behavior is influenced by the smaller ones. This agrees with the ansatz by Rege et al. \cite{bib17} concerning collapse of cells. For a given solid fraction, the PSD has a very slight influence on the specific energy density and permanent set.

\section{Conclusion}\label{sec4}
In this paper, the influence of different structural parameters, namely pore-size distribution (PSD), solid fraction and pore wall thickness, on the elastic and the inelastic behavior of cellular-like open-porous materials is investigated using a computational model. For a better understanding of the structure-property relation, one needs to look at the PSD rather than the probability density function. In conclusion, PSD and solid fraction are the most sensitive geometric properties which highly influences the bulk response of the structure. The solid fraction is sufficient to describe the linear elastic behavior of a porous material. It also highly influences the specific energy density and the residual deformation under cyclic loading. However, PSD plays an important role 
in capturing the complete description of the stress-strain response, including elastic, plateau and densification regimes. PSD and solid fraction are dimensionless parameters, and  independent on the scale (macro-scale or nano-scale). The dimensional parameters, such as cell wall diameter and the size of representative volume element are less significant, as they are interrelated to given PSD and solid fraction and depend on the scale in which the pore width intervals are defined. On the other hand, one needs to account for both, the PSD and the solid fraction, while correlating the synthesis and morphological parameters of the porous materials for synthesizing materials tailored for a specific application, as they are primary geometric properties that influence the bulk mechanical response.

%
%

%

\end{document}